\documentclass{article}

\usepackage{PRIMEarxiv}

\usepackage[utf8]{inputenc} % allow utf-8 input
\usepackage[T1]{fontenc}    % use 8-bit T1 fonts
\usepackage{hyperref}       % hyperlinks
\usepackage{url}            % simple URL typesetting
\usepackage{booktabs}       % professional-quality tables
\usepackage{amsfonts}       % blackboard math symbols
\usepackage{nicefrac}       % compact symbols for 1/2, etc.
\usepackage{microtype}      % microtypography
\usepackage{lipsum}
\usepackage{graphicx}
\graphicspath{{media/}}     % organize your images and other figures under media/ folder
\usepackage{subcaption}
\usepackage{verbatim}
\usepackage{float}
\usepackage{amsmath}
\usepackage{amssymb} 
\usepackage[round]{natbib}  % natbib with author-year citations, round parentheses

\hypersetup{
    pdftitle  = {PAUSE: A Privacy-Preserving Self-Reflection Tool for AI-Associated Cognitive Offloading},
    pdfauthor = {Mahbub Ul Alam, mahbub.ul.alam.anondo@gmail.com},
    pdfsubject  = {PAUSE: A Privacy-Preserving Self-Reflection Tool for AI-Associated Cognitive Offloading},
    pdfkeywords = {cognitive offloading, generative AI, AI literacy, human-AI interaction, self-reflection, critical thinking, creativity, privacy by design, reflective systems},
    colorlinks = true,
    linkcolor  = blue,
    urlcolor   = blue,
    citecolor  = blue
}

\usepackage{qrcode}      % Generates QR codes
\usepackage{orcidlink}   % Official ORCID icon + link

\title{

PAUSE: A Privacy-Preserving Self-Reflection Tool for AI-Associated Cognitive Offloading
\thanks{\textbf{PAUSE} (\textbf{P}atterns of \textbf{A}I \textbf{U}se: \textbf{S}elf-\textbf{E}xamination) is a public self-reflection tool for AI usage patterns.}}

\author{
  Mahbub Ul Alam \orcidlink{0000-0002-1101-3793}   \\
  SciLifeLab Data Centre, Uppsala University, Sweden \\
  \texttt{mahbub.ul.alam@scilifelab.uu.se, mahbub.ul.alam.anondo@gmail.com} \\ 
}

\begin{document}
\maketitle

\begin{abstract}

\begin{center}
    \qrcode[height=1.5cm]{https://anondo1969.github.io/pause}
    \vspace{.2cm}
    \small $\Longleftarrow$ \textit{Click or scan to open PAUSE}
\end{center}

Cognitive offloading is the use of external aids, such as notes, calculators, or search engines, to reduce mental effort. Large language models (LLMs) extend this to thinking itself, and by AI-associated cognitive offloading, I mean the pattern where a person routinely substitutes LLM output for their own reasoning, idea generation, learning, or communication. Recent empirical work reports associations between some patterns of LLM use and changes in critical thinking effort, neural engagement during assisted tasks, creative diversity, learning behaviour, and social dependence. Validated instruments for AI reliance, dependence, and literacy have begun to appear. I describe \textbf{PAUSE (Patterns of AI Use: Self-Examination)}\footnote{\textbf{PAUSE tool:} \href{https://anondo1969.github.io/pause}{anondo1969.github.io/pause}}, a privacy-by-design web tool (\textbf{link:} \href{https://anondo1969.github.io/pause}{anondo1969.github.io/pause}) that occupies a different niche from these. It is a lightweight, non-diagnostic reflection aid for private individual use. PAUSE is organised around how a person's own LLM use may relate to cognitive offloading across four everyday domains (\textit{reasoning \& critical thinking}, \textit{creativity \& originality}, \textit{research \& learning}, and \textit{social \& communicative capacity}). It delivers a short, free, no-login self-check, scores it entirely in the browser, and returns descriptive, domain-aware reflections. The self-check pairs reverse-scored behavioural items with a claim-evaluation reasoning probe, an alternative-uses creativity probe, and a small retrospective before-and-after block. PAUSE does not assume that AI use is harmful. It addresses the narrower case where AI substitutes for effort a person may want to preserve. The application is privacy-preserving by design:\footnote{\textbf{PAUSE source code:} \href{https://github.com/anondo1969/pause}{github.com/anondo1969/pause}} scoring is deterministic and runs client-side, no personal data is required, nothing is transmitted or stored beyond the browser session, and no LLM is involved in production. In this paper, I describe the conceptual grounding, the item design, the scoring, the architecture, and the design philosophy.

\textbf{PAUSE is a self-reflection tool, not a validated psychological instrument. It presents no reliability or validity evidence, and all readings should be interpreted as descriptive self-reflection rather than psychological measurement.}
\end{abstract}


\section{Introduction}\label{sec:introduction}

\begin{figure}[ht]
\centering
\includegraphics[width=\textwidth]{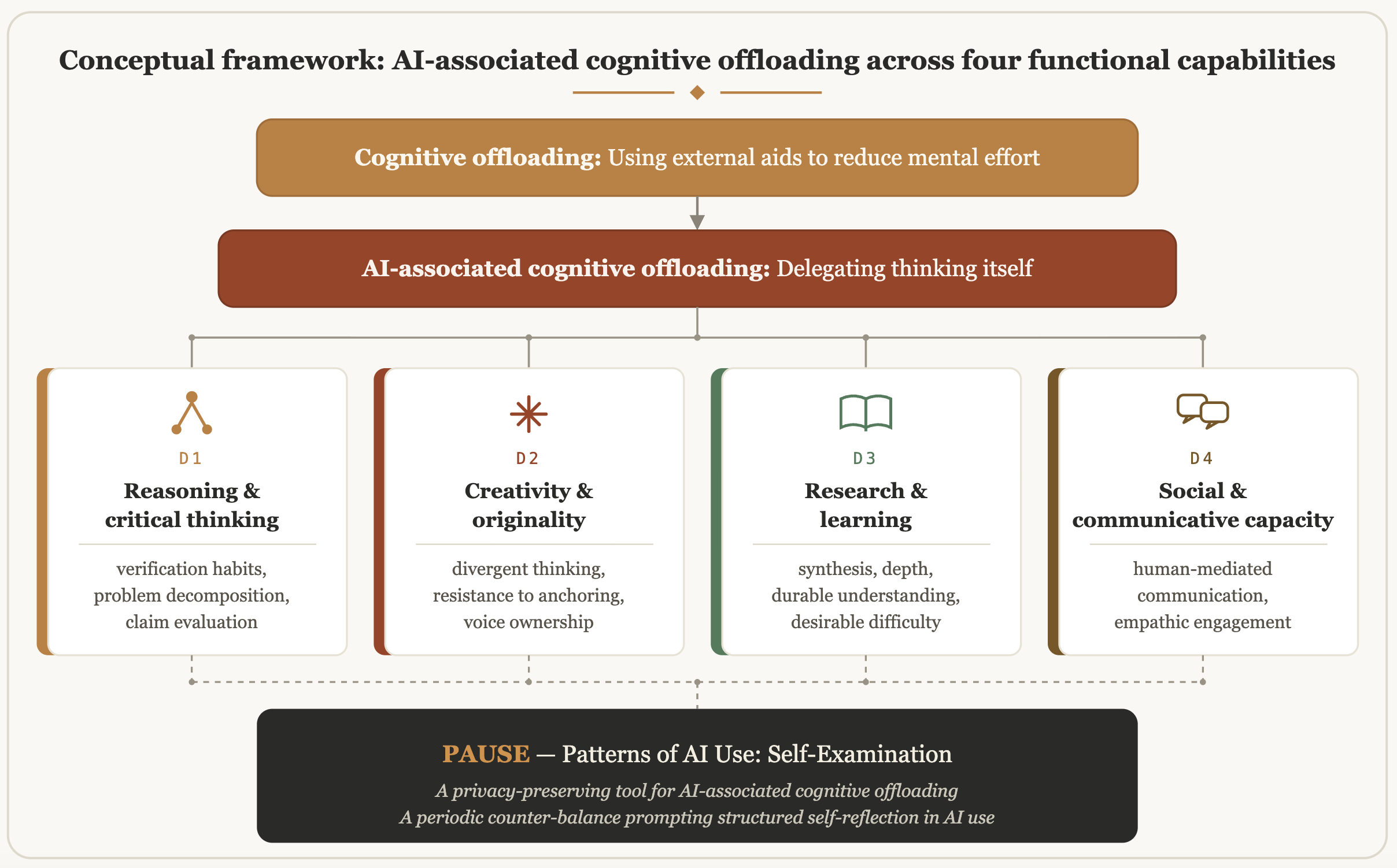}
\caption{Conceptual framework: cognitive offloading across four functional capabilities. The four domains D1 through D4 map onto distinct strands of the 2023--2026 empirical literature (Section~\ref{sec:empirical}) and feed into the self-check (Section~\ref{sec:four-dimensions}).}
\label{fig:fig1}
\end{figure}

In four years, generative artificial intelligence (AI) has moved from a research curiosity to a daily cognitive aid for hundreds of millions of people. The benefits are real and well documented: faster writing, broader idea generation, more accessible expertise. What remained unclear until recently was the cost of this cognitive aid.

That cost is best captured as AI-associated cognitive offloading. Cognitive offloading is the use of external action or aids to reduce the mental demand of a task \citep{risko2016cognitive}. Writing a note, reaching for a calculator, and running a search are all familiar forms of it. By AI-associated cognitive offloading I mean the form this habit takes with large language models (LLMs), where what gets delegated is the thinking itself: reasoning through a problem, generating ideas, learning new material, or communicating with other people.

A growing empirical literature has recently begun to address these consequences. Across an EEG-instrumented writing study, surveys of knowledge workers, creativity experiments, and a month-long randomised trial of chatbot use, one pattern recurs. The more a person substitutes LLM output for their own cognitive effort, the more measurable the shift on the faculty being substituted \citep{kosmyna2025brain, gerlich2025a, lee2025critical, doshi2024generative, holzner2025generative, fang2025chatbot}. This holds for neural engagement, enacted critical thinking, idea diversity, and loneliness and social dependence. The association appears to persist even when users understand intellectually that the LLM is not sentient \citep{kosmyna2025brain, lee2025critical}. Awareness alone does not appear to be protective, and the effects seem to track habitual use. Section~\ref{sec:empirical} reviews this literature in detail.

Existing critical thinking scales such as Watson-Glaser and the California Critical Thinking Skills Test (CCTST) are paid, generic, and pre-LLM. The General Attitudes towards Artificial Intelligence Scale (GAAIS) of \citet{schepman2020gaais} measures attitudes. AI literacy scales measure knowledge \citep{wang2023measuring, ng2021ailiteracy}. A person who has read about \textit{cognitive debt} \citep{kosmyna2025brain}, de-skilling, or over-reliance has no straightforward, public, and citation-grounded way to ask how any of it applies to them.

To address this, I built a privacy-by-design web tool that acts as a periodic counter-balance: each use inserts a moment of structured self-reflection into a person's relationship with their AI tools. Its name, \textbf{PAUSE} (\textbf{\textit{P}}\textit{atterns of} \textbf{\textit{A}}\textit{I} \textbf{\textit{U}}\textit{se:} \textbf{\textit{S}}\textit{elf-}\textbf{\textit{E}}\textit{xamination}), signals the same intent. PAUSE does not measure habit strength, underlying ability, or neural change. It surfaces self-reported behavioural patterns that the literature associates with the concerns above, and because it is cross-sectional, it describes current offloading patterns rather than accumulated \textit{cognitive debt}. Figure~\ref{fig:fig1} shows how the four PAUSE domains map onto the literature. Figure~\ref{fig:item-anatomy} shows what \textit{citation-grounding} means at the item level.

Two clarifications are worth making. Cognitive offloading is not inherently harmful; humans have long delegated thinking to notes, calculators, maps, calendars, search engines, and other people, usually with good results. PAUSE also does not assume AI use is harmful. Its concern is narrower: it targets unreflective, substitution-heavy offloading, where AI takes over effort a person may want to retain, and it is designed to prompt reflection rather than deliver a verdict.

In summary, PAUSE is designed to make a person briefly stop using software, attend to their own thinking, and receive evidence-based, domain-aware suggestions for more deliberate engagement. The most defensible reading of a PAUSE result is:

\begin{quote}
\textit{``In the last two weeks, my self-reported behaviour around AI use matches a pattern that recent empirical literature associates with cognitive offloading across four everyday domains (reasoning \& critical thinking, creativity \& originality, research \& learning, and social \& communicative capacity). The literature does not establish that I am suffering, declining, or harmed. It establishes that this pattern is associated, in studies of larger groups, with measurable changes in the relevant capability. The point of this tool is to make my own pattern visible to me, so that I can choose differently if I want to.''}
\end{quote}

\subsection{What this paper means by ``AI''}\label{sec:scoping}

Throughout this paper, ``AI'' refers to consumer-facing generative artificial intelligence built on large language models (LLMs): chat assistants such as ChatGPT, Claude, Gemini, and Copilot, their API (Application Programming Interface)-accessible models, and the smart-reply, autocomplete, and summarisation features built on top of them. This is the class of tool whose cognitive consequences the 2023--2026 literature documents, and it is the class the items ask the respondent to introspect about. The terms ``AI'', ``LLM'', ``GenAI'', ``chatbot'', and ``AI tool'' are used interchangeably below.

The self-check does not cover classical recommender systems, pre-LLM predictive features such as spam filters, or embodied and task-specific AI. I scope it this way for an empirical reason. The studies anchoring the four domains (Section~\ref{sec:four-dimensions}) concern LLM-class tools used in cognitive tasks, and PAUSE inherits that scope. The respondent-facing wording uses ``AI'' for accessibility, and the landing page spells out the scope for anyone who wants the specifics.

\subsection{Contributions}

I make three contributions in this paper:

\begin{enumerate}
    \item A four-domain mapping from the 2023--2026 empirical literature on LLM-associated cognitive effects to a self-reflection structure, with item-level citations (Appendix~\ref{app:items}).
    \item A public, citation-grounded set of reflective items, reverse-scored to counter agreement bias, paired with two short behavioural probes (a claim-evaluation reasoning probe and an Alternative Uses creativity probe) and a non-scored retrospective before-and-after block (Appendix~\ref{app:items} and Appendix~\ref{app:rubrics}).
    \item A privacy-by-design, no-backend reference web application that performs all scoring client-side, collects and transmits nothing, and involves no LLM in production; deployable as a static site or self-hosted (\textbf{link:} \href{https://anondo1969.github.io/pause}{anondo1969.github.io/pause}).
\end{enumerate}

\section{Related Work}\label{sec:related}

\subsection{Cognitive offloading and the LLM era}

PAUSE builds on the cognitive offloading framework of \citet{risko2016cognitive}. \citet{singh2025protecting} frame the influence as bidirectional. AI mimics aspects of human reasoning and, in doing so, reshapes the cognition of its users, particularly in the problem-framing phase that precedes consultation. \citet{shaw2026surrender} sharpen the vocabulary by separating \textit{cognitive offloading} (delegating a discrete task while one's own reasoning stays engaged) from \textit{cognitive surrender} (adopting an AI output with minimal scrutiny, so that agency itself transfers). Their framing is about calibration, and PAUSE's concern sits at the offloading end of this spectrum, where substitution is quiet and reflective control is what is at stake.

\citet{riley2025humanai} survey the emerging evidence through the psychological triad of cognition, behaviour, and emotion, structured around Bloom's cognitive levels and the I-PACE (Interaction of Person, Affect, Cognition, Execution) behavioural framework \citep{brand2016ipace}, and identify three behavioural trajectories: positive (AI scaffolds cognition), negative (AI displaces agency, degrading skills and inviting overreliance), and mixed. PAUSE's four domains map most directly onto the negative trajectory, while its recommendations draw on the positive one (Section~\ref{sec:counterdesign-rw}). The emotional strand of the triad (anticipatory anxiety, AI guilt, emotional dependence) is acknowledged as related but falls outside PAUSE's current scope: the self-check avoids items that measure affective states as such, and where the social domain touches emotionally salient situations, its items are framed as communicative substitution rather than as measures of emotional dependence.

\subsection{Empirical findings across four functional domains}\label{sec:empirical}

\textbf{Reasoning \& critical thinking (D1).} \citet{kosmyna2025brain}, in an electroencephalography (EEG) instrumented essay-writing study at the MIT Media Lab ($n = 54$; preprint), found that participants using ChatGPT showed the weakest neural connectivity across memory, attention, and executive networks, the lowest sense of ownership over their work, and impaired recall of their own writing, with lingering effects when they later wrote unaided; the authors coined \textit{cognitive debt} for this reduced neural engagement. \citet{gerlich2025a}, in a mixed-methods study of 666 participants, found a significant negative correlation between AI tool use and critical thinking, mediated by cognitive offloading and strongest in younger users. \citet{lee2025critical}, surveying 319 knowledge workers, reported that higher confidence in generative AI predicts less enacted critical thinking, and that knowledge work shifts from problem-solving toward verifying and integrating AI output. \citet{dergaa2024tools} name the mechanism AI-Chatbot-Induced Cognitive Atrophy: the gradual weakening of judgment through habitual uncritical acceptance of chatbot output. The offloaded faculty here is evaluation itself, so its erosion is the hardest kind for the user to detect.

\textbf{Creativity \& originality (D2).} \citet{doshi2024generative} found that LLM access boosts individual story creativity, especially for less inherently creative writers, while significantly homogenising output across writers. \citet{anderson2024homogenization} and the meta-analysis of \citet{holzner2025generative} confirm the diversity reduction across divergent thinking, idea generation, and real-world artefacts. \citet{kumar2024creativity} found that exposure to LLM-generated frameworks reduced the diversity of participants' idea sets both during and after use, suggesting lasting fixation. \citet{wadinambiarachchi2024effects} observed the same fixation in novice designers using Midjourney, whose designs showed less variety than those made with a simple image search or no aid, and who frequently copied visual elements from the AI images. \citet{bastani2025educational} and \citet{dellacqua2023jagged} provide complementary field evidence.

\textbf{Research \& learning (D3).} \citet{abbas2024harmful} found that university students who used ChatGPT heavily reported growing difficulty retaining knowledge. \citet{etkin2025differential} showed that readers who already understood a topic performed worse on comprehension tests when given AI summaries instead of the original text; the summaries stripped the contextual detail skilled readers depend on. \citet{peters2025generalization} demonstrated that language models systematically drop qualifiers when summarising research, presenting findings as more general than the evidence supports. \citet{tankelevitch2024metacognitive} identify forming goals, forming queries, and judging output as the loci where critical thinking must be deliberately maintained. On learning outcomes, \citet{bastani2025educational} found that high-school students given GPT-4 access solved more problems in the moment but performed worse than the control group once the tool was removed. \citet{liu2026persistence} provide causal evidence for the same pattern in randomised controlled trials ($N=1{,}222$), adding that AI assistance also reduced persistence. At population scale, \citet{rismanchian2026faster}, analysing $3.2$ million adaptive-tutor interactions over ten years, found that study time on AI-susceptible problems fell sharply after ChatGPT's release, with a durable cost on later proctored retention items.

\textbf{Social \& communicative capacity (D4).} \citet{fang2025chatbot}, in a four-week randomised controlled trial with 981 participants and over 300,000 messages, found dose-dependent increases in loneliness and emotional dependence with chatbot use, alongside lower socialisation with real people. \citet{hohenstein2023ai} demonstrated that AI-drafted email replies became more polished and uniformly positive, yet recipients who suspected AI involvement trusted the sender less. The efficiency gain carried a relational cost. \citet{jakesch2023human} showed that even expert evaluators cannot reliably distinguish AI-written from human-written text, and \citet{riley2025humanai} note the downstream effect. When any message could plausibly be AI-generated, the authenticity signal of personal writing is diluted for everyone.

\subsection{Counter-design and structured prompting}\label{sec:counterdesign-rw}

Not all evidence points toward decline. \citet{gerlich2025b}, the provocations study of \citet{drosos2025itmakesthinkprovocations}, and the diverse-persona work of \citet{wan2025diverse} all suggest that \textbf{how} a user engages with AI matters more than \textbf{whether}. Structured prompting, deliberate verification, alternation between AI-assisted and AI-free work, and exposure to diverse outputs can mitigate or reverse offloading effects. The session-four result of \citet{kosmyna2025brain}, in which brain-only participants subsequently using ChatGPT outperformed sustained LLM users, supports this directly. \citet{makransky2025wow} showed that a generative-AI tutoring chatbot which prompted students to connect ideas and explain their reasoning produced better assessment performance than traditional instruction. In the same spirit, \citet{hong2025cognitive} found that instruction which deliberately offloaded lower-order writing tasks to AI, freeing attention for analysis and revision, produced larger critical-thinking gains. Offloading, designed well, can serve the very capacities it is often assumed to erode. I draw on this counter-design literature, along with the harm literature, when designing the recommendations.

\subsection{Existing tools}

Several validated instruments occupy adjacent constructs: the AI Dependence Scale (AIDep-22) of \citet{wu2026aidep}, whose four factors include a cognitive-dependence dimension built on the offloading framework; the SAID questionnaire of \citet{garciacastro2025said} for AI dependence in secondary-school students; and, for AI \textit{literacy}, the performance-based GLAT of \citet{jin2024glat}, alongside the older GAAIS \citep{schepman2020gaais} and AI literacy scales \citep{wang2023measuring, ng2021ailiteracy, carolus2023mails}. Consumer-facing sites also host ad hoc ``AI dependency'' quizzes without methodology or citations.

Closest in spirit is the \textit{Offloading Score} of \citet{padmakumar2026offloading}, which is likewise framed both for individual reflection and for designers reducing overreliance. It estimates the fraction of cognitive effort offloaded by reconstructing a counterfactual workflow from behavioural logs (validated with $n=40$ developers on programming tasks). PAUSE instead asks a descriptive, reverse-scored questionnaire organised by four everyday domains, and stores nothing.

Against this landscape, I position PAUSE as a \textbf{reflective design probe}. The measurement corners (reliance, dependence, literacy) are already occupied by validated scales, and PAUSE borrows from that literature rather than trying to out-measure it. What is distinctive is the combination: descriptive and non-diagnostic, organised by four functional domains, anchored item by item to the literature, and privacy-preserving to the point of storing nothing. \citet{riley2025humanai} call for grounded evaluation frameworks that balance benefits against emerging human-centric risks; PAUSE offers one deliberately modest, public, and deployable answer to that call.

\section{Design of the Self-Check}\label{sec:instrument}

\subsection{Domains}\label{sec:four-dimensions}

I organise reflection around four domains. These are a working vocabulary grounded in the literature, not a tested factor structure.

\begin{enumerate}
    \item \textbf{D1: Reasoning \& Critical Thinking.} The habit and capacity to decompose problems, evaluate evidence, verify claims, and form independent conclusions.
    \item \textbf{D2: Creativity \& Originality.} The habit and capacity to generate novel, divergent, and personally voiced ideas not anchored on AI templates.
    \item \textbf{D3: Research \& Learning.} The habit and capacity to engage with material at depth, to synthesise across sources, and to retain understanding rather than the AI-summarised gist.
    \item \textbf{D4: Social \& Communicative Capacity.} The habit and capacity to draft, negotiate, and connect with other humans without AI mediation.
\end{enumerate}

Each domain is treated as a continuum. Self-reported behavioural patterns are read as indicators of offloading depth, not as direct measures of underlying ability. In the ``jagged frontier'' sense of \citet{dellacqua2023jagged}, AI may help on some tasks within a domain and hurt on others, and a domain-level mean compresses that variation.

\subsection{Item design principles}

I developed the items under the following constraints:

\begin{figure}[ht]
  \centering
  \includegraphics[width=\linewidth]{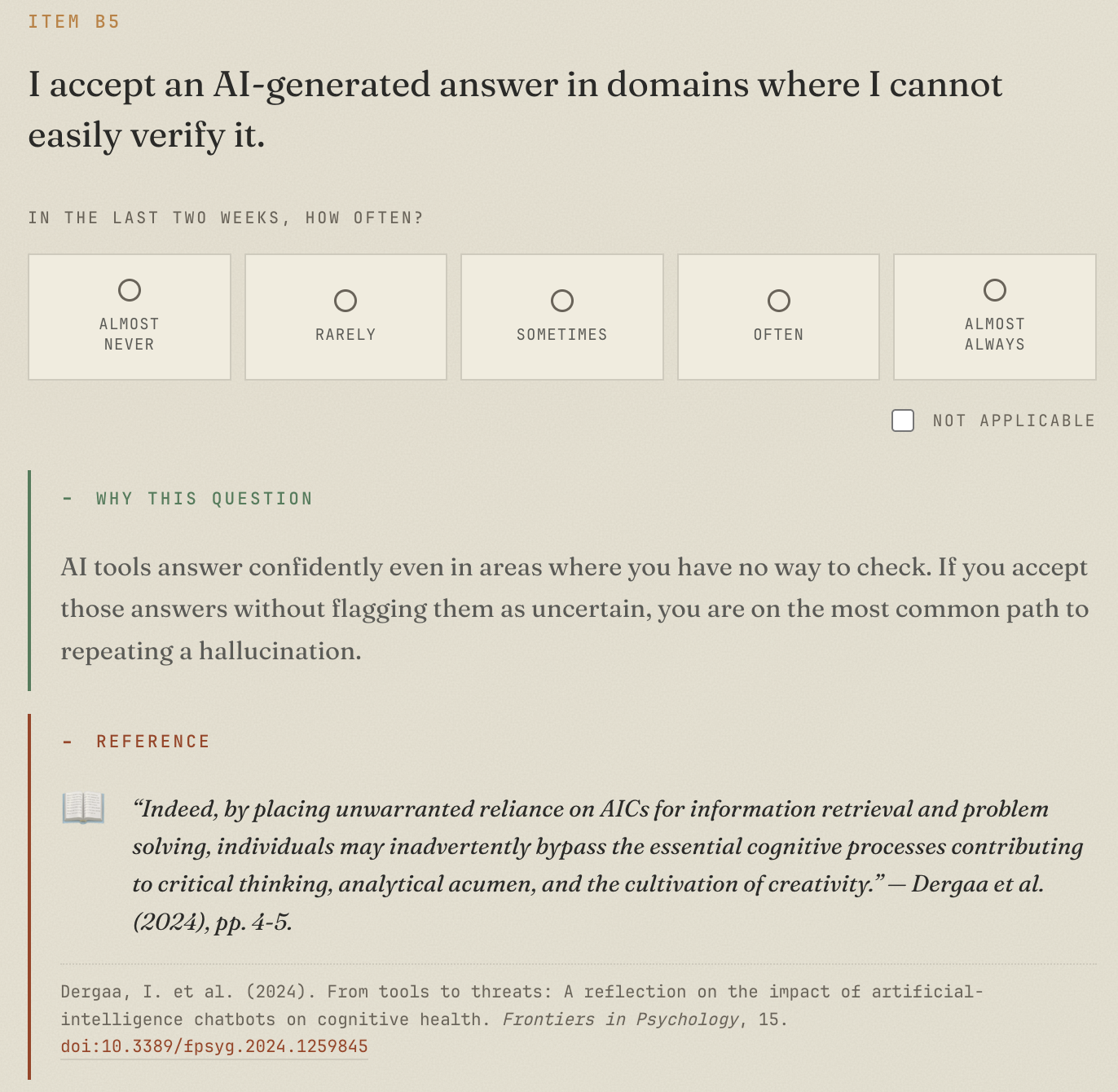}
  \caption{What \textit{citation-grounded} means, shown on item B5 as rendered by the deployed tool. Every scored item has the same three-part anatomy: the behavioural task with its response scale and a \emph{Not applicable} option; an expandable \emph{Why this question} panel giving the item's rationale in plain language; and an expandable \emph{Reference} panel containing an exact quote from the anchoring study (here \citet{dergaa2024tools}, pp.~4--5) above the full citation with a resolvable DOI. Both panels can be opened before answering, so the respondent can always inspect the evidence behind a question rather than take the item on trust. The full item-to-anchor mapping is in Appendix~\ref{app:items}.}
  \label{fig:item-anatomy}
\end{figure}

\begin{itemize}
    \item Every item targeting LLM-era behaviour is anchored in a peer-reviewed or arXiv paper from 2023 to 2026 (Figure~\ref{fig:item-anatomy}); only BL0 (I-PACE baseline-stability framing, \citealt{brand2016ipace}) and C7 (Alternative Uses Task, \citealt{guilford1967nature}) use older canonical references. The full mapping is in Appendix~\ref{app:items}.
    \item Items are behavioural (``\textit{In the last two weeks, how often did you \ldots}'') rather than dispositional (``\textit{I am the kind of person who \ldots}''), to reduce social-desirability bias.
    \item Roughly one third of the items per domain (two or three of seven) are reverse-scored to counter acquiescence; the exact set is marked in Appendix~\ref{app:items}.
    \item Two domains (D1 and D2) include a short behavioural probe to add a limited behavioural texture to self-report.
    \item All items have a ``\textit{not applicable}'' option for users whose work does not involve a domain.
\end{itemize}

\subsection{Scoring}\label{sec:scoring}

For each domain $d \in \{\text{D1}, \text{D2}, \text{D3}, \text{D4}\}$:

\begin{itemize}
    \item Each Likert item is scored 0 to 4 (with reverse coding applied).
    \item Probe items are scored 0 to 4 against pre-specified rubrics (Section~\ref{sec:probes}; full rubrics in Appendix~\ref{app:rubrics}).
    \item The domain reading $D_d$ is the mean of valid item scores, scaled to 0 to 100. Higher values indicate greater self-reported offloading. They do not indicate cognitive impairment.
    \item A domain reading uses only the items the respondent actually answered; N/A responses are dropped, not imputed. If fewer than three items in a domain were answered, the tool shows no number for that domain and marks it \textit{limited basis}. A domain answered only in part carries a short note that the reading rests on fewer items.
    \item Each domain reading is mapped to one of four descriptive bands for display: \textit{Minimal offloading signal} (0 to 24), \textit{Mild} (25 to 49), \textit{Moderate} (50 to 74), and \textit{Strong offloading signal} (75 to 100). These are fixed, quarter-width cut points on the 0 to 100 scale, chosen for legibility rather than derived from a norming sample, and they carry no diagnostic meaning.
\end{itemize}

\textbf{There is no overall composite.} A single number across four domains of unequal weight and uneven character would compress more than it reveals, on the jagged-frontier grounds above (\citet{dellacqua2023jagged}), so the tool neither computes nor shows one. The four domains are reported separately.

\subsection{Domain branching}

Before scoring, the user selects a primary self-identification: \textit{Researcher or academic}, \textit{Creative professional}, \textit{Student}, \textit{Knowledge worker}, \textit{Software or data engineer}, or \textit{Other}. All users answer the same items, but the feedback recommendations branch on this selection. A high D1 reading for a researcher emphasises peer review, replication-checking, and brain-only literature reading sessions; the same reading for a creative professional emphasises journal-based ideation, blank-page warmups, and AI-free first drafts.

\subsection{Behavioural probes}\label{sec:probes}

To add a behavioural texture beyond self-report, I embed two short probes:

\begin{itemize}
    \item \textbf{Reasoning probe (B7).} A claim-evaluation multiple-choice item. The respondent reads a short paragraph reporting a research-like claim and selects the central methodological flaw from four options, one of which is ``the conclusion seems sound.'' Scoring is 0 (correct flaw identified), 2 (a non-central flaw identified), or 4 (uncritical acceptance).
    \item \textbf{Creativity probe (C7).} A 90-second Alternative Uses Task (AUT; \citealt{guilford1967nature}) on a common object. Responses are scored for fluency (count of distinct entries) and basic flexibility (rough category diversity) using deterministic client-side heuristics. The scoring thresholds are provisional and are disclosed as such on the result page.
\end{itemize}

Both probes are answerable only while their timer runs. A respondent may also skip either probe explicitly, and a B7 timer that expires without an answer is treated the same way: in both cases the probe is scored as \emph{not applicable} and dropped from the domain mean, exactly as a Likert item marked N/A. Non-response is never scored as option~(d); running out of time is a different behaviour from uncritically accepting a claim, and conflating the two would bias the reading upward for slow readers and interrupted respondents.

Both probes draw from stratified pools with client-side, cross-session rotation, so a returning respondent does not see the same fallacy type or object category twice in a row. Rotation state travels only in the optional return token (Section~\ref{sec:returntoken}), which carries no respondent-identifying information, and no probe content is transmitted anywhere. Pools, selection logic, and full rubrics are in Appendix~\ref{app:rubrics}.

\subsection{Retrospective before-and-after block}

After the context section and before D1, I add a small retrospective block framed: \textit{``Compared to before you regularly used AI tools, how often do you now\ldots''}. Four behavioural items (BL1-BL4) map to the four domains: independent fact-checking, idea generation from scratch, reading original sources end-to-end, and drafting personal messages on your own. The response scale runs from \textit{much less often} to \textit{much more often}.

These items are \textbf{not} aggregated into the domain readings. They are reported separately as a \textbf{trajectory indicator}. A purely cross-sectional reading cannot distinguish a user who \textit{never} engaged in a given habit from one who \textit{used to} and has stopped, and retrospective self-report is noisy and biased, so the block does not solve that problem, but it surfaces the distinction. An optional opening item (BL0) asks whether pre-AI habits in these areas were already stable; a low answer earns a note that the before-and-after direction should be read as suggestive. The block is skippable for respondents with under \textbf{twelve months} of regular LLM use, whose pre-AI baseline was likely sporadic rather than stable.

\section{PAUSE web application}\label{sec:implementation}

\subsection{Architecture}

I built PAUSE as a static multi-page web application with no backend. The frontend is plain HTML, CSS, and JavaScript, with no build step, no framework runtime, and no third-party scripts or requests of any kind; the typefaces are self-hosted under the SIL Open Font License, so loading a page contacts only the site's own host. All scoring and result rendering run in the user's browser. No server endpoint accepts responses, and the application layer logs nothing. The site is served via GitHub Pages directly from the source repository, so the running tool and its code sit in one place, and it is equally self-hostable on any static host or as a container image. No LLM is involved in production. That choice is deliberate. A tool that helps a person reflect on cognitive offloading should not itself offload.

\subsection{Privacy by design}\label{sec:returntoken}

\begin{itemize}
    \item No account, no upload, and no personally identifiable information collected by the application.
    \item Responses, scores, and free-text inputs never leave the user's browser and are never logged.
    \item The only client-side storage is session-scoped and cleared on tab close, and it holds nothing but two short rotation codes recording which probe variants were seen last, so a repeat session does not present the same fallacy type or object category (Section~\ref{sec:probes}). Responses and scores are held in memory only while the page is open; they are never written to storage, and a page refresh discards them.
    \item Return-visit comparison works through a self-contained scorecard token. The result page shows a short opaque string encoding the current domain readings, which the user copies and stores themselves. On a return visit, pasting it displays the earlier readings alongside the current ones. The token is never transmitted, never persisted by the application, and never linked to any account or identity. When tokens from different scoring versions are pasted together, the comparison view flags that readings across versions may not be directly comparable; the instrument version is also displayed in the site footer.
    \item Any platform-level analytics are governed by the static host's own policy. The PAUSE application itself records nothing about what the respondent answered.
\end{itemize}

\subsection{Accessibility}

The self-check works on mobile and desktop browsers, with keyboard
navigation, screen-reader-friendly markup (semantic HTML, ARIA labels on form controls, countdown timers excluded from live-region announcements), and a colour palette whose text pairings meet the WCAG~2.1~AA contrast threshold. All items sit at a plain-English reading level. The application requires JavaScript by construction: rendering and scoring are performed client-side precisely so that nothing is transmitted, and respondents with JavaScript disabled see a static notice explaining this. The tool ships in English; Swedish and Bengali translations are planned.

\subsection{Result framing and ethical safeguards}

I frame PAUSE throughout as a tool for self-reflection rather than a measurement instrument. A persistent banner states that it is not validated, that a reading need not reflect any stable underlying trait, and that its readings should not drive real-world decisions. In keeping with this, the four band labels (\textit{Minimal}, \textit{Mild}, \textit{Moderate}, and \textit{Strong offloading signal}) are marked as descriptive rather than normed (defined in Section~\ref{sec:scoring}).

The result page avoids clinical language: no respondent is described as ``impaired'' or ``at risk''. Its header names the reading in neutral terms of AI-use patterns. A high reading is neither a verdict nor a reason to stop using AI tools, and a low reading is no kind of clearance: each reports only the match between recent self-reported behaviour and a pattern the literature associates with offloading. My aim is conscious, structured engagement, not abstinence.

I present the domain-aware recommendations as practices to consider, drawn from the counter-design literature (Section~\ref{sec:counterdesign-rw}), not as prescriptions, treatment advice, or claims tested on the individual respondent. A reading is for that respondent's own reflection and must not justify any consequential decision about them or anyone else, whether hiring, admissions, performance evaluation, or clinical or educational assessment.

PAUSE is not designed for minors. A soft, non-blocking age check precedes the self-check; a negative response explains that the tool is not intended for those under 18 and invites the respondent to return when they are. Because the design is privacy-preserving, I deliberately avoid a hard age gate that would require identity verification. Nothing a respondent enters is collected, transmitted, or stored beyond the session, so the attestation is retained nowhere and no respondent's identity is captured at any point.

\section{Worked example: a single result and a six-month comparison}\label{worked-example}

The following is illustrative. The respondent is fabricated and the numbers are constructed to show the mechanics; no real user data is involved.

\subsection{A single result}

A creative professional (a freelance writer) selects \textit{Creative professional}, reports 6--15 hours per week with AI tools, and over two years of regular use. Their D1 (\textit{Reasoning \& Critical Thinking}) items score as shown in Table~\ref{tab:worked-d1}, with the three reverse-coded items flipped so a higher coded value always means more offloading.

\begin{table}[ht]
  \centering
   \begin{tabular}{llcc}
\hline
Item & Response & & Coded \\
\hline
B1 (rev.) & Checks AI facts against a source & Almost never & 4 \\
B2        & Reaches for AI before framing     & Often        & 3 \\
B3 (rev.) & Pushes back on plausible AI reasoning & Sometimes & 2 \\
B4        & Harder to hold an argument in their head & Often     & 3 \\
B5        & Accepts AI answers they cannot verify & Often       & 3 \\
B6 (rev.) & Articulates disagreement first    & Rarely       & 3 \\
B7 (probe)& Identifies a non-central flaw      & ---          & 2 \\
\hline
\multicolumn{3}{l}{Total across the seven scored items}       & 20 \\
%\hline
\end{tabular}
\caption{Worked D1 (\textit{Reasoning \& Critical Thinking}) scoring for the illustrative respondent of Section~\ref{worked-example}. The three reverse-scored items (B1, B3, B6) are shown \emph{after} flipping, so a higher coded value always indicates more self-reported offloading. The seven-item total of 20 gives a mean of $20/7 = 2.86$, which scales to a domain reading of 71, a \emph{Moderate offloading signal} under the fixed bands of Section~\ref{sec:scoring}.}
  \label{tab:worked-d1}
\end{table}

The mean is $20/7 = 2.86$, scaled to $\mathbf{71}$ on 0--100: a \textit{Moderate offloading signal} (Section~\ref{sec:scoring}). Their full result is \textit{Reasoning \& Critical Thinking} 71, \textit{Creativity \& Originality} 93, \textit{Research \& Learning} 71, \textit{Social \& Communicative Capacity} 75, that is, \textit{Moderate}, \textit{Strong}, \textit{Moderate}, and \textit{Strong offloading signal}. Each domain appears as a labelled bar with one practice:

\begin{itemize}
  \item \textit{Reasoning \& Critical Thinking}, \textsc{moderate offloading signal 71}: ``Before opening an AI tool, spend 10 minutes writing the problem in your own words. Keep that page.''
  \item \textit{Creativity \& Originality}, \textsc{strong offloading signal 93} (with \textsc{C7 uses provisional thresholds ($\pm$1 band)}): ``Try a one-week AI-free draft. Just one draft. Notice what changes in your voice.''
  \item \textit{Research \& Learning}, \textsc{moderate offloading signal 71}: ``When learning a new
  technique or reference, work from the original source for an hour before turning to AI explainers.''
  \item \textit{Social \& Communicative Capacity}, \textsc{strong offloading signal 75}: ``Send one personal, AI-free message to a friend or collaborator this week. Notice if you feel exposed.''
\end{itemize}

A short profile-aware summary sits above the bars:

\begin{quote}
As a creative professional, the areas that matter most are likely \textit{Creativity \& Originality} (keeping your own voice, generating your own first ideas) and \textit{Reasoning \& Critical Thinking} (pushing back on AI suggestions that sound plausible but flatten your work). Your scores suggest a high level of offloading across most areas. This does not mean you are doing something wrong, and this tool is not a diagnostic. But the pattern is pronounced enough that a deliberate experiment, one AI-free working session this week in whichever area matters most to you, may be genuinely revealing. The point is not to stop using AI; it is to find out what your own capabilities feel like without it, right now, so you can make a conscious choice.
\end{quote}

\begin{table}[ht]
  \centering
\begin{tabular}{lcccc}
\hline
Domain & Feb & Jul & Change & Band \\
\hline
Reasoning \& Critical Thinking  & 64 & 41 & $-23$ & Moderate $\rightarrow$ Mild \\
Creativity \& Originality & 88 & 62 & $-26$ & Strong $\rightarrow$ Moderate \\
Research \& Learning   & 52 & 18 & $-34$ & Moderate $\rightarrow$ Minimal \\
Social \& Communicative Capacity     & 35 & 80 & $+45$ & Mild $\rightarrow$ Strong \\
\hline
\end{tabular}
  \caption{The synthetic six-month demo trajectory (February to July)
  shipped with the Compare view. The four domains trace deliberately
  distinct paths rather than parallel drift: three decline at different rates while Social rises throughout and ends in a different band from where it started, and the final readings span all four descriptive bands. \emph{Change} is the last-minus-first difference on the 0 to 100 scale; band transitions use the fixed cut points of Section~\ref{sec:scoring}. The data are fabricated for illustration and involve no real respondent.}
  \label{tab:demo-trajectory}
\end{table}

\subsection{A six-month comparison}

Comparison over time uses the return token (Section~\ref{sec:returntoken}): the respondent copies a short opaque string at each session and keeps it; pasting several back reconstructs the trajectory entirely in the browser, and PAUSE stores nothing between visits. The tool ships with a synthetic demo trajectory of six monthly sessions (February to July), summarised in Table~\ref{tab:demo-trajectory}. 

\begin{figure}[p]
  \centering
  \begin{subfigure}{0.92\linewidth}
    \centering
    \includegraphics[width=\linewidth]{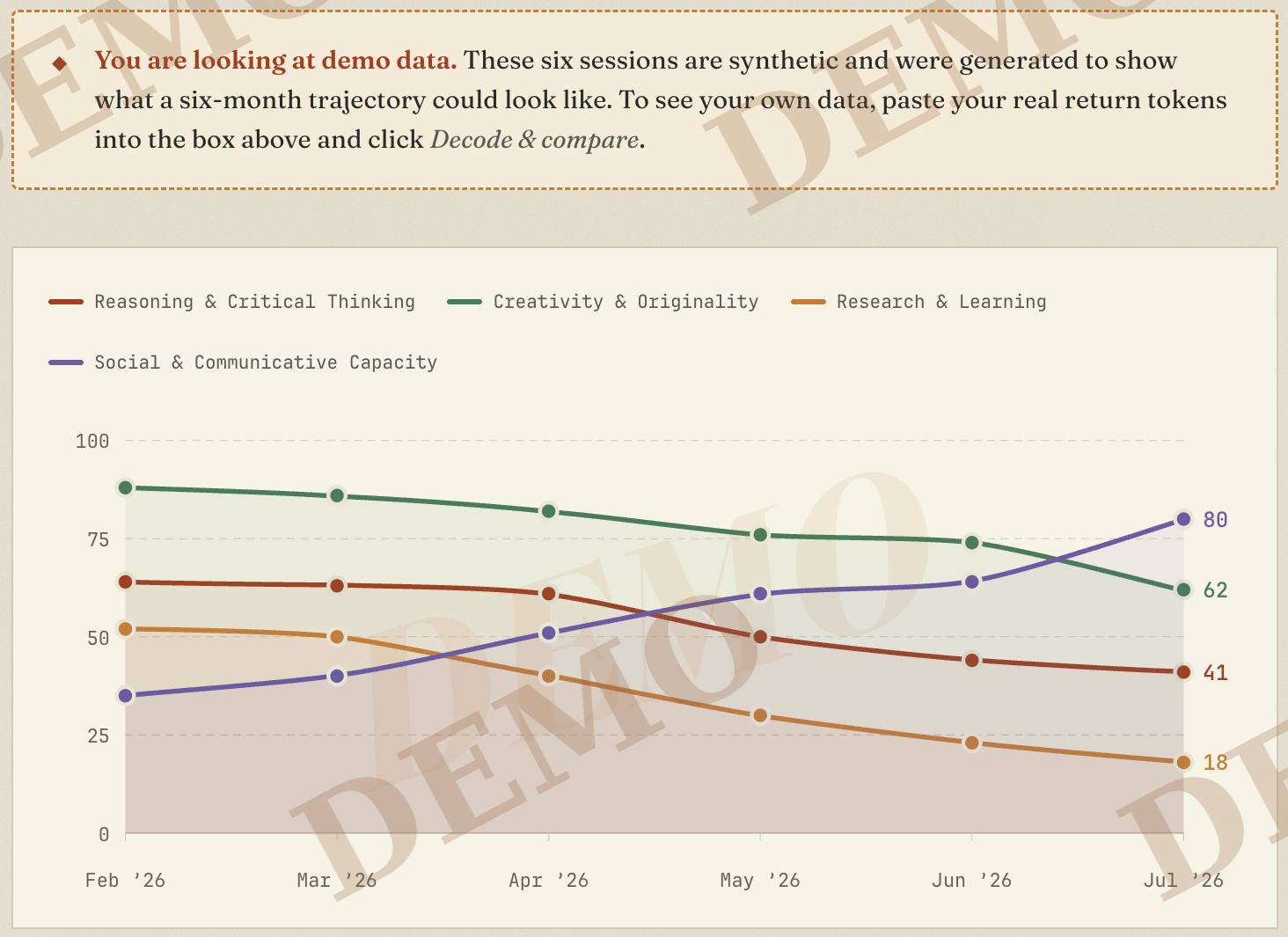}
    \caption{The trajectory chart: one smoothed line per domain over six monthly sessions, with final readings labelled at the line ends. The diagonal watermark marks the data as synthetic even in a cropped screenshot.}
    \label{fig:compare-chart}
  \end{subfigure}

  \vspace{1.1em}

  \begin{subfigure}{0.92\linewidth}
    \centering
    \includegraphics[width=\linewidth]{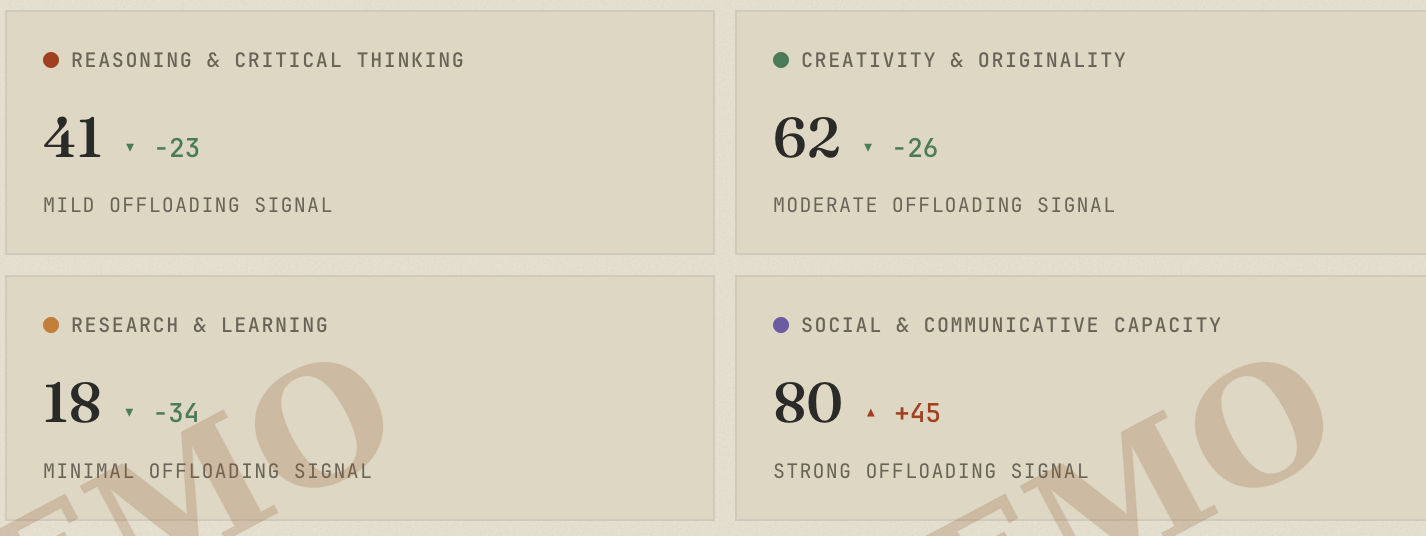}
    \caption{Per-domain delta cards: the latest reading, the last-minus-first change on the 0 to 100 scale, and the descriptive band of the latest session.}
    \label{fig:compare-deltas}
  \end{subfigure}

  \caption{The Compare view rendering the synthetic six-month demo
  trajectory of Table~\ref{tab:demo-trajectory}, shown top to bottom as the respondent sees it. Everything in all these panels is computed in the browser from the pasted return tokens alone; no token is transmitted, stored, or linked to any identity.}
  \label{fig:compare-demo}
\end{figure}

\begin{figure}[ht]
  \centering
  \includegraphics[width=\linewidth]{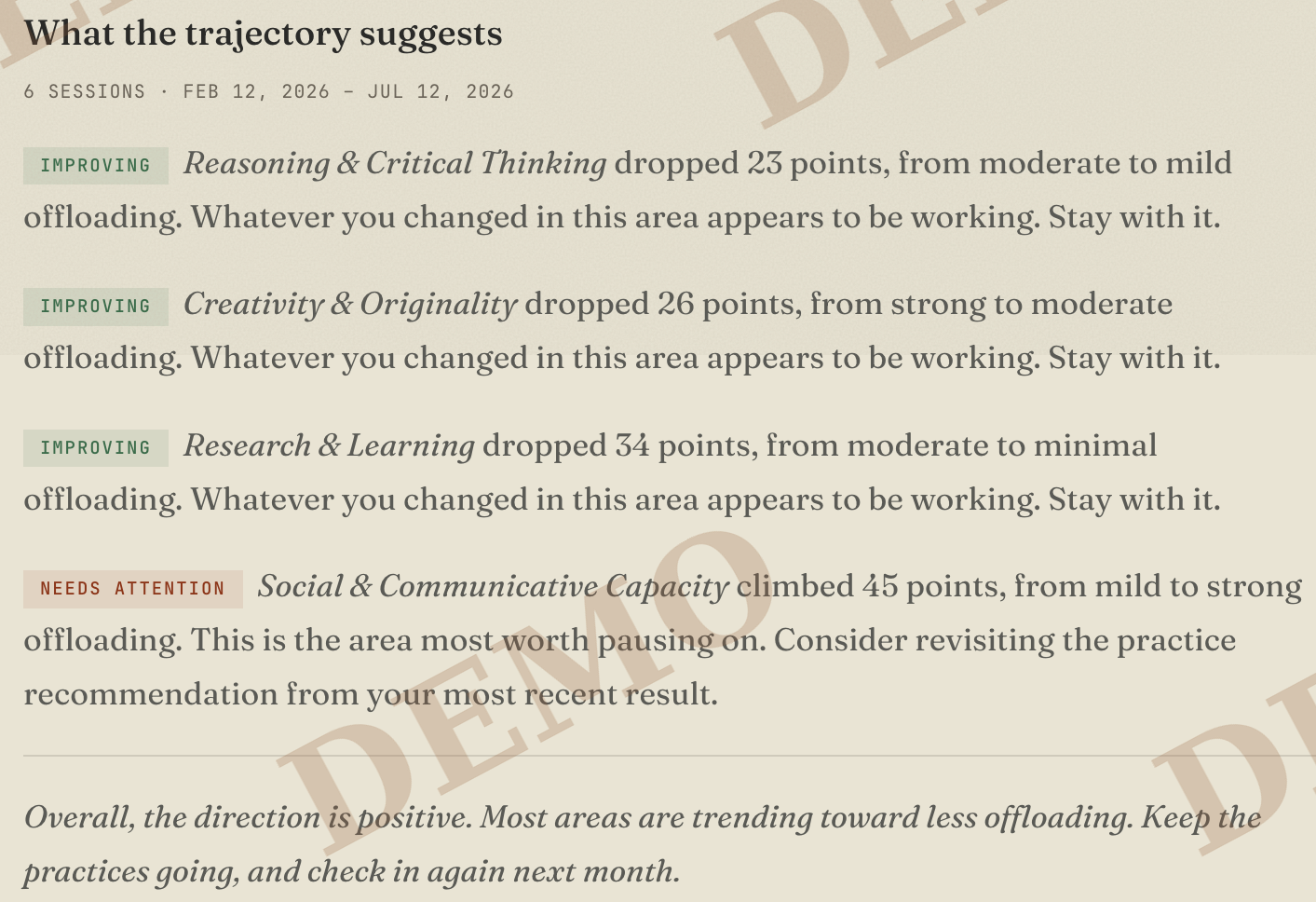}
  \caption{The trajectory panel narrates each change in the same neutral register.}
  \label{fig:narrated-analysis}
\end{figure}

Figure~\ref{fig:compare-chart} shows the trajectory chart and
Figure~\ref{fig:compare-deltas} the per-domain delta cards, as the
respondent sees them; the narrated analysis panel that follows is provided in Figure~\ref{fig:narrated-analysis}.

\section{Discussion}\label{sec:discussion}

\subsection{Implications for design, education, and policy}

For AI tool designers, D1 and D3 name behaviours that current chat interfaces tend to encourage: friction-free acceptance, summarisation in place of source reading, and immediate answer delivery. The provocations literature indicates that small interface choices, such as asking the user to commit to a position before showing output, surfacing source links by default, or pacing the response, can shift these behaviours \citep{drosos2025itmakesthinkprovocations, gerlich2025b, makransky2025wow}. PAUSE supplies a vocabulary for naming what such choices target.

For educators and workplace learning teams, the domain readings can scaffold curriculum design around the specific behaviours associated with offloading rather than around general AI literacy. A high D2 reading in a creative writing programme calls for AI-free first drafts and journal-based ideation; a high D3 reading in a research methods course calls for source-reading exercises and supervised summary verification. The domain-branching feature is a first attempt to translate readings into role-specific suggestions of this kind.

For policy actors concerned with the cognitive effects of generative AI, the tool is a low-cost, publicly available reference that can be self-administered at population scale without infrastructure cost.

\subsection{Why a self-report tool?}

Self-report has known limitations (Section~\ref{sec:limitations}), and higher-fidelity methods exist: behavioural tasks under controlled conditions, neuroimaging, longitudinal performance tracking. I chose self-report deliberately, because it meets a use case those methods cannot, which is individual-level access at scale. A person curious about their own AI use cannot run an EEG study on themselves, and a teacher cannot administer a four-hour cognitive battery to a class. The value is in being short, free, and immediately interpretable, and the two embedded probes add a limited behavioural signal without giving up that brevity.

\subsection{Limitations}\label{sec:limitations}

\textbf{Self-report of offloading is susceptible to the very faculty it concerns.} A person whose critical-thinking habits have decayed may also have decayed insight into the decay. The behavioural probes partially mitigate this without eliminating it.

\textbf{Correlation is not causation.} The tool cannot show that a given user's pattern was \textit{caused by} their LLM use; it can only describe a pattern. For the same reason, a reading cannot establish that AI use changed anyone's abilities.

\textbf{Sampling is non-representative.} People who voluntarily take a self-check on AI and cognition are not a population sample.

\textbf{The four-domain structure is a working organisation,} not a confirmed factor structure. Whether ``AI-associated cognitive offloading'' is even a coherent construct, distinct from general technology dependence or media-multitasking effects, is an open question.

\textbf{Several anchoring studies are recent preprints.} The literature on LLM cognitive effects turns over faster than peer-review cycles. Because PAUSE's stance is descriptive, the bar for anchoring a reflective item is lower than for a causal claim, but anchors remain supporting evidence for an item's design, not proof that the item behaves as intended.

\textbf{Several items additionally conflate early consultation with substitution:} a respondent who deliberately brings AI in at the start of a task as a scaffold (the engagement pattern Section~\ref{sec:counterdesign-rw} describes favourably) will honestly endorse items such as B2, C1, and D1 and read as offloading. The items measure when and how often AI enters the workflow, which is an imperfect proxy for whether one's own reasoning stayed engaged.

\textbf{The retrospective block's directional labels should likewise be read as self-narrative rather than measurement}; the interface presents them separately from the domain readings and with an explicit stability caveat for that reason.

\textbf{The probes are simple.} B7 is multiple-choice and therefore easier than free-response flaw identification, and respondents trained in epidemiology, statistics, or experimental design will systematically read as less offloading, which is partly the construct and partly a training effect. The pool is at risk of memorisation once public, mitigated by the client-side rotation. The C7 heuristic captures fluency and category-level flexibility but not originality or elaboration; \citet{silvia2008assessing} describe higher-validity alternatives that the no-LLM-in-production constraint precludes, and the result page carries an explicit band-uncertainty disclosure.

\section{Conclusion}

I designed PAUSE to make visible what the 2023--2026 empirical literature has begun to document. Habitual substitution of LLM output for one's own cognitive effort is associated with detectable shifts in reasoning, creativity, learning, and social engagement, and PAUSE turns that literature into a structured self-reflection protocol delivered through privacy-preserving software.

The design reflects three deliberate choices. First, epistemic humility. Readings describe self-reported offloading patterns, there is no overall composite, and nothing is diagnosed. Second, transparency. Every item is anchored to a specific empirical finding, and the full item set and scoring code are open from launch. Third, privacy by design. The tool records none of the respondent's answers, sends nothing to a server, and runs no LLM in production.

The limits matter as much as the features: the tool is a reflection aid, its probes are simple, some of its anchors are preprints, and its four domains are a working organisation. If PAUSE serves its purpose, people will take it briefly and periodically, notice their patterns, and choose differently where they want to. It is my attempt at public, citation-grounded, privacy-preserving self-reflection for one of the defining cognitive shifts of this decade.

\section{Availability}\label{sec:availability}

\begin{itemize}
    \item \textbf{Source code:} \href{https://github.com/anondo1969/pause}{github.com/anondo1969/pause} (MIT licence)
    
    \item \textbf{Self-check items and scoring:} released as Appendix~\ref{app:items} and Appendix~\ref{app:rubrics} of this paper and as a standalone CC-BY 4.0 document
    
    \item \textbf{Live tool:} \href{https://anondo1969.github.io/pause}{anondo1969.github.io/pause} (GitHub Pages), no login required, no data collected
    
    \item \textbf{Verification suite:} a reproducible \textit{Node.js} test harness in the repository re-derives the Section~\ref{worked-example} worked example from the production scoring module (the D1 reading of 71 and the brick AUT example, band~1), asserts the Section~\ref{worked-example} band rubric and the Section~\ref{sec:probes} cross-session rotation guarantee over 2{,}000 draws, fuzzes the return-token decoder, verifies HTML escaping and the WCAG~2.1~AA contrast of the deployed palette, confirms the absence of third-party requests, and drives the complete self-check and comparison flows in a simulated browser, including the probe skip paths. Every mechanically checkable claim in this paper is checked by it; psychometric properties are not, and are not claimed.
    
\end{itemize}

The scoring module, the AUT category dictionary, the B7 pool, and the feedback library are all inspectable source files in the public repository.

\appendix

\section{Full Item List}\label{app:items}

All items have a ``Not applicable'' option in addition to the response scale shown. For the Likert items this is a checkbox; for the two timed probes (B7, C7) it takes the form of an explicit skip control, and a B7 timer that expires without an answer is likewise scored as not applicable (Section~\ref{sec:probes}). Each behavioural item is anchored to a study that motivated its design; anchors are supporting evidence for an item's framing, not validation of it.

\textit{Sections A, BL, B, C, D, and E correspond to entries in the deployed \texttt{items.json} (the questionnaire flow). The age check (H) is implemented as a pre-self-check modal.}

\subsection*{Section A: Context (not part of the reading)}

\begin{itemize}
    \item \textbf{A1.} Which best describes your primary work? \textit{Researcher / Creative / Student / Knowledge worker / Software engineer / Other.}
    \item \textbf{A2.} Average hours per week with AI tools? \textit{$< 1$ / 1-5 / 6-15 / 16-30 / $> 30$.}
    \item \textbf{A3.} How long have you used LLM-based AI tools regularly? \textit{$< 6$ months / 6-12 months / 1-2 years / $> 2$ years.}
\end{itemize}

\subsection*{Section BL: Retrospective trajectory (reported separately, skippable if A3 $< 12$ months)}

\begin{itemize}
    \item \textbf{BL0.} \textit{Before I started using AI tools regularly, my work and study habits in these four areas (verification of facts, generating ideas, reading sources, drafting messages) were already well-established and stable.} (0-4 agreement) \textit{Anchor: \citet{brand2016ipace}.}
    \item \textbf{BL1.} Compared to before you regularly used AI tools, how often do you now \textbf{verify factual claims against independent sources}? ($-2$ to $+2$) \textit{Anchor: \citet{gerlich2025a}.}
    \item \textbf{BL2.} Compared to before, how often do you now \textbf{generate ideas from scratch without external input} when starting creative work? ($-2$ to $+2$) \textit{Anchor: \citet{kumar2024creativity}.}
    \item \textbf{BL3.} Compared to before, how often do you now \textbf{read original sources end to end} rather than relying on summaries? ($-2$ to $+2$) \textit{Anchor: \citet{etkin2025differential}.}
    \item \textbf{BL4.} Compared to before, how often do you now \textbf{draft personal or sensitive messages entirely on your own}? ($-2$ to $+2$) \textit{Anchor: \citet{hohenstein2023ai}.}
\end{itemize}

\subsection*{Section B: D1; Reasoning \& Critical Thinking (B1, B3, B6 reverse-scored; B7 probe)}

Likert (B1-B6): \textit{Almost never (0) / Rarely (1) / Sometimes (2) / Often (3) / Almost always (4)}

\begin{itemize}
    \item \textbf{B1.} When an AI gives me a factual answer that I will rely on (a statistic, a date, a citation), I check it against an independent source before using it. (Reverse) \textit{Anchor: \citet{lee2025critical}; \citet{tankelevitch2024metacognitive}.}
    \item \textbf{B2.} When I face a complex problem, I now reach for an AI tool before I have spent meaningful time framing the problem myself. \textit{Anchor: \citet{lee2025critical}; \citet{kosmyna2025brain}.}
    \item \textbf{B3.} When an AI's reasoning seems plausible, I push back on it and ask myself where it could be wrong. (Reverse) \textit{Anchor: \citet{drosos2025itmakesthinkprovocations}.}
    \item \textbf{B4.} I find it harder than I used to find it to hold a long argument together in my head without external help. \textit{Anchor: \citet{kosmyna2025brain}; \citet{gerlich2025a}.}
    \item \textbf{B5.} I accept an AI-generated answer in domains where I cannot easily verify it. \textit{Anchor: \citet{dergaa2024tools}.}
    \item \textbf{B6.} When I disagree with an AI's output, I take time to articulate why, in writing or aloud, before deciding what to do. (Reverse) \textit{Anchor: \citet{tankelevitch2024metacognitive}.}
    \item \textbf{B7. [Probe]} Claim-evaluation multiple-choice item drawn from a stratified pool (Appendix~\ref{app:rubrics}). \textit{Anchors: \citet{lee2025critical}; \citet{tankelevitch2024metacognitive}.}
\end{itemize}

\subsection*{Section C: D2; Creativity \& Originality (C2, C5 reverse-scored; C7 probe)}

\begin{itemize}
    \item \textbf{C1.} When I sit down to create something, I open an AI tool before I have produced any of my own ideas. \textit{Anchor: \citet{doshi2024generative}; \citet{kumar2024creativity}.}
    \item \textbf{C2.} I keep a private space (notebook, sketch app, raw doc) where I generate ideas without any AI in the loop. (Reverse) \textit{Anchor: \citet{kosmyna2025brain}; \citet{kumar2024creativity}.}
    \item \textbf{C3.} I notice my finished work sounds more like a generic AI register than like me. \textit{Anchor: \citet{holzner2025generative}; \citet{doshi2024generative}.}
    \item \textbf{C4.} When I get an AI suggestion I like, I tend to keep most of its structure rather than rewrite it in my own way. \textit{Anchor: \citet{doshi2024generative}; \citet{anderson2024homogenization}.}
    \item \textbf{C5.} In the last month, I have produced a piece of work I am proud of that contained no AI-generated content. (Reverse) \textit{Anchor: \citet{kosmyna2025brain}.}
    \item \textbf{C6.} When I'm stuck creatively, my first move is to ask an AI rather than to stay with the discomfort. \textit{Anchor: \citet{kumar2024creativity}.}
    \item \textbf{C7. [Probe]} 90-second Alternative Uses Task on a randomised common object. \textit{Anchor: \citet{guilford1967nature}.}
\end{itemize}

\subsection*{Section D: D3; Research \& Learning (D2, D5 reverse-scored)}

\begin{itemize}
    \item \textbf{D1.} When I need to learn something new, I ask an AI for a summary instead of reading the source. \textit{Anchor: \citet{abbas2024harmful}.}
    \item \textbf{D2.} I can paraphrase, in my own words and without notes, the main argument of the last serious piece I ``read'' through AI summarisation. (Reverse) \textit{Anchor: \citet{kosmyna2025brain}.}
    \item \textbf{D3.} I have caught myself citing or paraphrasing a claim that turned out to be a hallucination or misattribution. \textit{Anchor: \citet{peters2025generalization}.}
    \item \textbf{D4.} When I ``understand'' a topic via AI explanation, I later struggle to apply it to a new case. \textit{Anchor: \citet{bastani2025educational}.}
    \item \textbf{D5.} I do at least one weekly task in my domain (reading, coding, writing, analysis) deliberately without AI, to keep the skill warm. (Reverse) \textit{Anchor: \citet{kosmyna2025brain}.}
    \item \textbf{D6.} I skip the slow, frustrating part of learning (struggle, dead ends, false starts) because AI lets me. \textit{Anchor: \citet{kumar2024creativity}; \citet{abbas2024harmful}.}
    \item \textbf{D7.} My notes, drafts, and intermediate work increasingly consist of pasted AI output rather than my own writing. \textit{Anchor: \citet{lee2025critical}.}
\end{itemize}

\subsection*{Section E: D4; Social \& Communicative Capacity (E3, E7 reverse-scored)}

\begin{itemize}
    \item \textbf{E1.} I draft personal or sensitive messages (to friends, family, colleagues) by getting an AI to write them first. \textit{Anchor: \citet{hohenstein2023ai}.}
    \item \textbf{E2.} When something is difficult or personal, I find it easier to have an AI help me work through it than to talk to a person. \textit{Anchor: \citet{fang2025chatbot}.}
    \item \textbf{E3.} I have had a meaningful, AI-free conversation longer than 20 minutes in the last week (in person, by phone, or by video call). (Reverse) \textit{Anchor: \citet{fang2025chatbot}.}
    \item \textbf{E4.} I notice myself reading other people's writing and assuming it must be AI-generated. \textit{Anchor: \citet{jakesch2023human}; \citet{riley2025humanai}.}
    \item \textbf{E5.} I prefer AI responses to human ones because the AI is faster, more available, or less judgmental. \textit{Anchor: \citet{fang2025chatbot}.}
    \item \textbf{E6.} I have reduced the time I spend in human-to-human casual exchange (in person, on the phone, in messaging) since I started using AI tools heavily. \textit{Anchor: \citet{fang2025chatbot}.}
    \item \textbf{E7.} When I face a small social or interpersonal decision, I work it out by thinking it through myself or talking to a trusted person before, or instead of, asking an AI. (Reverse) \textit{Anchor: \citet{tankelevitch2024metacognitive}.}
\end{itemize}

\subsection*{Section H: Age check}

\begin{itemize}
    \item \textbf{H1.} Are you 18 years of age or older? (yes / no)
\end{itemize}

\section{Probe pools and rubrics}\label{app:rubrics}

\subsection*{B7 claim-evaluation pool}

The B7 pool contains items stratified across 10 reasoning-fallacy types, with several items per type. Each item presents a short paragraph with a research-like claim about AI productivity, learning, or use, followed by four options that ask the respondent to identify the central methodological flaw. The correct flaw is one specific option (varied across items so the correct letter is not predictable); one option is always ``the conclusion seems sound to me.''

\paragraph{Fallacy taxonomy.} The 10 types and their two-letter codes are: \texttt{ss} small sample size; \texttt{sb} selection bias / unrepresentative sample; \texttt{cf} confounding / correlation-causation; \texttt{rm} regression to the mean; \texttt{cb} confirmation bias / non-independent confirmation; \texttt{es} effect size vs.\ statistical significance; \texttt{br} base rate neglect; \texttt{sv} survivorship bias; \texttt{an} anchoring / framing effect; \texttt{ge} hasty generalisation / single-case extrapolation.

\paragraph{Item selection at presentation.} Item selection runs entirely client-side, in two stages: (1) determine the set of allowed fallacy types (all 10 unless the respondent has an avoidance hint from a prior session, in which case the most recently seen type is excluded); (2) pick a fallacy type uniformly at random from the allowed set, then pick an item uniformly at random from that type's items.

\paragraph{Cross-session avoidance.} The avoidance hint is communicated between sessions via the optional return token (Section~\ref{sec:returntoken}). The token carries a two-letter fallacy-type code and a three-letter C7 category code; these are the only non-score fields, and neither carries respondent-identifying information. The fallacy type never repeats across two consecutive sessions, and the chance of seeing a familiar paragraph is low even under perfect recall.

\paragraph{Scoring.} 0 = correct flaw identified; 2 = a non-(d) wrong flaw identified; 4 = option (d), ``the conclusion seems sound.''

\paragraph{Full pool.} Full text of all items, with fallacy-type codes, is in \texttt{data/instrument/b7\_pool.json} in the project repository. The file is a JSON object whose \texttt{pool} field is an array of items; each item has \texttt{id}, \texttt{fallacy\_type}, \texttt{paragraph}, \texttt{options} (array of four), \texttt{correct}, and \texttt{explanation\_after}.

\subsection*{C7 AUT scoring heuristic (deterministic, client-side)}

\paragraph{Object pool.} The C7 pool contains 12 common objects stratified across 6 everyday-object categories, with two objects per category:

\begin{itemize}
    \item \texttt{bld} (building material): brick, cinderblock
    \item \texttt{fst} (fastener / small office item): paperclip, rubber band
    \item \texttt{wrn} (wearable / personal item): shoe, leather belt
    \item \texttt{shl} (shelter / canopy): umbrella, tarpaulin
    \item \texttt{ctr} (container / vessel): plastic bucket, mason jar
    \item \texttt{txt} (textile / soft furnishing): bath towel, wool blanket
\end{itemize}

Object selection uses the same two-stage stratified picker as B7, with cross-session avoidance keyed on the category code rather than the fallacy type. This serves both as anti-memorisation and as protection against any object-specific affordance bias dominating a respondent's trajectory.

\paragraph{Fluency.} Count of distinct entries after trimming, lower-casing, stripping non-alphanumeric characters, and collapsing internal whitespace. Two entries that normalise to the same string are treated as duplicates and counted once.

\paragraph{Flexibility.} Each entry is mapped to one or more category tokens via a per-object dictionary (one dictionary per object, with roughly 50--80 keyword variants distributed across 10 categories each, bundled in \texttt{data/instrument/c7\_dictionaries.json}). Flexibility is the count of distinct categories spanned by the entries.

\paragraph{Combined score (0-4).}
\begin{itemize}
  \item $\geq 8$ fluent and $\geq 5$ categories $\rightarrow$ \textbf{0}: strong fluency and category diversity
  \item $\geq 5$ fluent and $\geq 3$ categories (not qualifying for band~0) $\rightarrow$ \textbf{1}: adequate fluency and diversity
  \item 3--4 fluent $\rightarrow$ \textbf{2}: moderate fluency, limited diversity
  \item 1--2 fluent $\rightarrow$ \textbf{3}: low fluency
  \item 0 fluent, or $\geq 5$ fluent with $< 3$ categories $\rightarrow$ \textbf{4}: very low fluency or all near-duplicates
\end{itemize}

Operationally, the band-4 ``all near-duplicates'' case fires when a respondent produces five or more entries that all map to one or two categories (e.g., five different ways to throw or break a brick all map to \texttt{weapon}); this is the intended ``fluent but fixated'' case. Fluency 3--4 with no category spread falls into band 2 rather than band 4 because the small entry count makes the fixation diagnosis unreliable.

\paragraph{Worked example.} For the prompt \emph{a brick}, the response

\begin{quote}
\small
\texttt{1. build a wall} \\
\texttt{2. use as a doorstop} \\
\texttt{3. sharpen a knife on it} \\
\texttt{4. hold down papers} \\
\texttt{5. as a step}
\end{quote}

\noindent normalises to five distinct entries (fluency $= 5$). Mapped to the brick dictionary: \texttt{build a wall} $\rightarrow$ building; \texttt{doorstop} $\rightarrow$ weight; \texttt{sharpen} $\rightarrow$ tool; \texttt{hold down papers} $\rightarrow$ weight; \texttt{step} $\rightarrow$ building. Distinct categories: \{building, weight, tool\}, flexibility $= 3$. Score: fluency 5 with flexibility 3 $\rightarrow$ band 1.

\paragraph{Threshold provenance.} The threshold values (8, 5, 5, 3, 3, 4, 2, 1) are provisional and were set ex ante based on the distribution of fluency and flexibility values reported in the AUT literature for adult samples on common-object prompts. Any revision would change the numerical cut-points without changing the band count or the band direction.

\paragraph{What the scoring heuristic does not capture.} The deterministic client-side scoring captures fluency and category-level flexibility but not originality (statistical rarity of a use) or elaboration (depth of description); within this domain, originality is carried by the self-report items rather than the probe. Originality scoring from the probe requires sample-level frequency data, which the tool does not collect; elaboration scoring requires human raters. Both are out of scope for the probe. Because the object pool and category dictionaries are public, the heuristic is also satisfiable by keyword stuffing; a respondent determined to reach band~0 can do so mechanically. The tool's threat model is self-deception rather than adversarial input, so this is disclosed rather than defended against.

% Bibliography
\bibliographystyle{apalike}
\bibliography{references}

@article{brand2016ipace,
  author    = {Brand, Matthias and Young, Kimberly S. and Laier, Christian and W{\"o}lfling, Klaus and Potenza, Marc N.},
  title     = {Integrating Psychological and Neurobiological Considerations Regarding the Development and Maintenance of Specific {I}nternet-Use Disorders: {A}n {I}nteraction of {P}erson-{A}ffect-{C}ognition-{E}xecution ({I-PACE}) Model},
  journal   = {Neuroscience \& Biobehavioral Reviews},
  volume    = {71},
  pages     = {252--266},
  year      = {2016},
  doi       = {10.1016/j.neubiorev.2016.08.033},
  url       = {https://doi.org/10.1016/j.neubiorev.2016.08.033},
  note      = {\href{https://doi.org/10.1016/j.neubiorev.2016.08.033}{doi:10.1016/j.neubiorev.2016.08.033}}
}

@article{bastani2025educational,
author = {Hamsa Bastani  and Osbert Bastani  and Alp Sungu  and Haosen Ge  and Özge Kabakcı  and Rei Mariman },
title = {Generative AI without guardrails can harm learning: Evidence from high school mathematics},
journal = {Proceedings of the National Academy of Sciences},
volume = {122},
number = {26},
pages = {e2422633122},
year = {2025},
doi = {10.1073/pnas.2422633122},
URL = {https://www.pnas.org/doi/abs/10.1073/pnas.2422633122},
eprint = {https://www.pnas.org/doi/pdf/10.1073/pnas.2422633122},
  note      = {\href{https://doi.org/10.1073/pnas.2422633122}{doi:10.1073/pnas.2422633122}}
}

@ARTICLE{dergaa2024tools,
    
AUTHOR={Dergaa, Ismail  and Ben Saad, Helmi  and Glenn, Jordan M.  and Amamou, Badii  and Ben Aissa, Mohamed  and Guelmami, Noomen  and Fekih-Romdhane, Feten  and Chamari, Karim },
           
TITLE={From tools to threats: a reflection on the impact of artificial-intelligence chatbots on cognitive health},
          
JOURNAL={Frontiers in Psychology},
          
VOLUME={15},
  
YEAR={2024},
  
URL={https://www.frontiersin.org/journals/psychology/articles/10.3389/fpsyg.2024.1259845},
  
DOI={10.3389/fpsyg.2024.1259845},
  
ISSN={1664-1078},
  note      = {\href{https://doi.org/10.3389/fpsyg.2024.1259845}{doi:10.3389/fpsyg.2024.1259845}}
}

@article{
doshi2024generative,
author = {Anil R. Doshi  and Oliver P. Hauser },
title = {Generative AI enhances individual creativity but reduces the collective diversity of novel content},
journal = {Science Advances},
volume = {10},
number = {28},
pages = {eadn5290},
year = {2024},
doi = {10.1126/sciadv.adn5290},
URL = {https://www.science.org/doi/abs/10.1126/sciadv.adn5290},
eprint = {https://www.science.org/doi/pdf/10.1126/sciadv.adn5290},
  note      = {\href{https://doi.org/10.1126/sciadv.adn5290}{doi:10.1126/sciadv.adn5290}}
}

@ARTICLE{etkin2025differential,
    
AUTHOR={Etkin, Hudson K.  and Etkin, Kai J.  and Carter, Ryan J.  and Rolle, Camarin E. },
           
TITLE={Differential effects of GPT-based tools on comprehension of standardized passages},
          
JOURNAL={Frontiers in Education},
          
VOLUME={10},
  
YEAR={2025},
  
URL={https://www.frontiersin.org/journals/education/articles/10.3389/feduc.2025.1506752},
  
DOI={10.3389/feduc.2025.1506752},
  
ISSN={2504-284X},
  note      = {\href{https://doi.org/10.3389/feduc.2025.1506752}{doi:10.3389/feduc.2025.1506752}}
}

@misc{fang2025chatbot,
      title={How AI and Human Behaviors Shape Psychosocial Effects of Extended Chatbot Use: A Longitudinal Randomized Controlled Study}, 
      author={Cathy Mengying Fang and Auren R. Liu and Valdemar Danry and Eunhae Lee and Samantha W. T. Chan and Pat Pataranutaporn and Pattie Maes and Jason Phang and Michael Lampe and Lama Ahmad and Sandhini Agarwal},
      year={2025},
      eprint={2503.17473},
      archivePrefix={arXiv},
      primaryClass={cs.HC},
      url={https://arxiv.org/abs/2503.17473},
      note      = {\href{https://doi.org/10.48550/arXiv.2503.17473}{doi:10.48550/arXiv.2503.17473}}
}

@article{gerlich2025a,
  author    = {Gerlich, Michael},
  title     = {{AI} Tools in Society: Impacts on Cognitive Offloading and the Future of Critical Thinking},
  journal   = {Societies},
  volume    = {15},
  number    = {1},
  pages     = {6},
  year      = {2025},
  doi       = {10.3390/soc15010006},
  url       = {https://doi.org/10.3390/soc15010006},
  note      = {\href{https://doi.org/10.3390/soc15010006}{doi:10.3390/soc15010006}}
}

@article{peters2025generalization,
  author    = {Peters, Uwe and Chin-Yee, Benjamin},
  title     = {Generalization Bias in Large Language Model Summarization of Scientific Research},
  journal   = {Royal Society Open Science},
  volume    = {12},
  pages     = {241776},
  year      = {2025},
  doi       = {10.1098/rsos.241776},
  url       = {https://doi.org/10.1098/rsos.241776},
  note      = {\href{https://doi.org/10.1098/rsos.241776}{doi:10.1098/rsos.241776}}
}

@article{risko2016cognitive,
  author    = {Risko, Evan F. and Gilbert, Sam J.},
  title     = {Cognitive Offloading},
  journal   = {Trends in Cognitive Sciences},
  volume    = {20},
  number    = {9},
  pages     = {676--688},
  year      = {2016},
  doi       = {10.1016/j.tics.2016.07.002},
  url       = {https://doi.org/10.1016/j.tics.2016.07.002},
  note      = {\href{https://doi.org/10.1016/j.tics.2016.07.002}{doi:10.1016/j.tics.2016.07.002}}
}

@article{abbas2024harmful,
  author    = {Abbas, Muhammad and Jam, Farooq Ahmed and Khan, Tariq Iqbal},
  title     = {Is It Harmful or Helpful? {E}xamining the Causes and Consequences of Generative {AI} Usage Among University Students},
  journal   = {International Journal of Educational Technology in Higher Education},
  volume    = {21},
  pages     = {10},
  year      = {2024},
  doi       = {10.1186/s41239-024-00444-7},
  url       = {https://doi.org/10.1186/s41239-024-00444-7},
  note      = {\href{https://doi.org/10.1186/s41239-024-00444-7}{doi:10.1186/s41239-024-00444-7}}
}

@article{hohenstein2023ai,
  author    = {Hohenstein, Jess and Kizilcec, Ren{\'e} F. and DiFranzo, Dominic and Aghajari, Zhila and Mieczkowski, Hannah and Levy, Karen and Naaman, Mor and Hancock, Jeffrey and Jung, Malte F.},
  title     = {Artificial Intelligence in Communication Impacts Language and Social Relationships},
  journal   = {Scientific Reports},
  volume    = {13},
  pages     = {5487},
  year      = {2023},
  doi       = {10.1038/s41598-023-30938-9},
  url       = {https://doi.org/10.1038/s41598-023-30938-9},
  note      = {\href{https://doi.org/10.1038/s41598-023-30938-9}{doi:10.1038/s41598-023-30938-9}}
}

@misc{kosmyna2025brain,
  author    = {Kosmyna, Nataliya and Hauptmann, Eugene and Yuan, Ye Tong and Situ, Jessica and Liao, Xian-Hao and Beresnitzky, Ashly Vivian and Braunstein, Iris and Maes, Pattie},
  title     = {Your Brain on {ChatGPT}: Accumulation of Cognitive Debt when Using an {AI} Assistant for Essay Writing Task},
  year      = {2025},
  eprint    = {2506.08872},
  archiveprefix = {arXiv},
  doi       = {10.48550/arXiv.2506.08872},
  url       = {https://arxiv.org/abs/2506.08872},
  note      = {\href{https://doi.org/10.48550/arXiv.2506.08872}{doi:10.48550/arXiv.2506.08872}}
}

@inproceedings{lee2025critical,
  author    = {Lee, Hao-Ping (Hank) and Sarkar, Advait and Tankelevitch, Lev and Drosos, Ian and Rintel, Sean and Banks, Richard and Wilson, Nicholas},
  title     = {The Impact of Generative {AI} on Critical Thinking: Self-Reported Reductions in Cognitive Effort and Confidence Effects from a Survey of Knowledge Workers},
  booktitle = {Proceedings of the 2025 CHI Conference on Human Factors in Computing Systems (CHI '25)},
  year      = {2025},
  publisher = {ACM},
  address   = {Yokohama, Japan},
  doi       = {10.1145/3706598.3713778},
  url       = {https://doi.org/10.1145/3706598.3713778},
  note      = {\href{https://doi.org/10.1145/3706598.3713778}{doi:10.1145/3706598.3713778}}
}

@article{makransky2025wow,
  author    = {Makransky, Guido and Shiwalia, Ban M. and Herlau, Tue and Blurton, Steven},
  title     = {Beyond the ``Wow'' Factor: Using Generative {AI} for Increasing Generative Sense-Making},
  journal   = {Educational Psychology Review},
  volume    = {37},
  pages     = {60},
  year      = {2025},
  doi       = {10.1007/s10648-025-10039-x},
  url       = {https://doi.org/10.1007/s10648-025-10039-x},
  note      = {\href{https://doi.org/10.1007/s10648-025-10039-x}{doi:10.1007/s10648-025-10039-x}}
}

@misc{riley2025humanai,
  author    = {Riley, Celeste and Al-Refai, Omar and Colunga Reyes, Yadira and Hammad, Eman},
  title     = {Human-{AI} Interactions: Cognitive, Behavioral, and Emotional Impacts},
  year      = {2025},
  eprint    = {2510.17753},
  archiveprefix = {arXiv},
  doi       = {10.48550/arXiv.2510.17753},
  url       = {https://arxiv.org/abs/2510.17753},
  note      = {\href{https://doi.org/10.48550/arXiv.2510.17753}{doi:10.48550/arXiv.2510.17753}}
}

@misc{singh2025protecting,
  author    = {Singh, Anjali and Taneja, Karan and Guan, Zhitong and Ghosh, Avijit},
  title     = {Protecting Human Cognition in the Age of {AI}},
  year      = {2025},
  eprint    = {2502.12447},
  archiveprefix = {arXiv},
  doi       = {10.48550/arXiv.2502.12447},
  url       = {https://arxiv.org/abs/2502.12447},
  note      = {\href{https://doi.org/10.48550/arXiv.2502.12447}{doi:10.48550/arXiv.2502.12447}}
}

@inproceedings{wadinambiarachchi2024effects,
  author    = {Wadinambiarachchi, Samangi and Kelly, Ryan M. and Pareek, Suranjana and Zhou, Qian and Velloso, Eduardo},
  title     = {The Effects of Generative {AI} on Design Fixation and Divergent Thinking},
  booktitle = {Proceedings of the 2024 CHI Conference on Human Factors in Computing Systems (CHI '24)},
  year      = {2024},
  publisher = {ACM},
  address   = {Honolulu, HI, USA},
  doi       = {10.1145/3613904.3642919},
  url       = {https://doi.org/10.1145/3613904.3642919},
  note      = {\href{https://doi.org/10.1145/3613904.3642919}{doi:10.1145/3613904.3642919}}
}

@inproceedings{anderson2024homogenization,
author = {Anderson, Barrett R and Shah, Jash Hemant and Kreminski, Max},
title = {Homogenization Effects of Large Language Models on Human Creative Ideation},
year = {2024},
isbn = {9798400704857},
publisher = {Association for Computing Machinery},
address = {New York, NY, USA},
url = {https://doi.org/10.1145/3635636.3656204},
doi = {10.1145/3635636.3656204},
booktitle = {Proceedings of the 16th Conference on Creativity \& Cognition},
pages = {413–425},
numpages = {13},
location = {Chicago, IL, USA},
series = {C\&C '24},
  note      = {\href{https://doi.org/10.1145/3635636.3656204}{doi:10.1145/3635636.3656204}}
}

@article{carolus2023mails,
title = {MAILS - Meta AI literacy scale: Development and testing of an AI literacy questionnaire based on well-founded competency models and psychological change- and meta-competencies},
journal = {Computers in Human Behavior: Artificial Humans},
volume = {1},
number = {2},
pages = {100014},
year = {2023},
issn = {2949-8821},
doi = {10.1016/j.chbah.2023.100014},
url = {https://www.sciencedirect.com/science/article/pii/S2949882123000142},
author = {Astrid Carolus and Martin J. Koch and Samantha Straka and Marc Erich Latoschik and Carolin Wienrich},
  note      = {\href{https://doi.org/10.1016/j.chbah.2023.100014}{doi:10.1016/j.chbah.2023.100014}}
}

@techreport{dellacqua2023jagged,
  author    = {Dell'Acqua, Fabrizio and McFowland III, Edward and Mollick, Ethan R. and Lifshitz-Assaf, Hila and Kellogg, Katherine and Rajendran, Saran and Krayer, Lisa and Candelon, Fran\c{c}ois and Lakhani, Karim R.},
  title     = {Navigating the Jagged Technological Frontier: Field Experimental Evidence of the Effects of {AI} on Knowledge Worker Productivity and Quality},
  institution = {Harvard Business School},
  year      = {2023},
  type      = {Working Paper},
  number    = {24-013},
  doi       = {10.2139/ssrn.4573321},
  url       = {https://ssrn.com/abstract=4573321},
  note      = {\href{https://doi.org/10.2139/ssrn.4573321}{doi:10.2139/ssrn.4573321}}
}

@article{ng2021ailiteracy,
author = {Ng, Davy Tsz Kit and Leung, Jac Ka Lok and Chu, Kai Wah Samuel and Qiao, Maggie Shen},
title = {AI Literacy: Definition, Teaching, Evaluation and Ethical Issues},
journal = {Proceedings of the Association for Information Science and Technology},
volume = {58},
number = {1},
pages = {504-509},
doi = {10.1002/pra2.487},
url = {https://asistdl.onlinelibrary.wiley.com/doi/abs/10.1002/pra2.487},
eprint = {https://asistdl.onlinelibrary.wiley.com/doi/pdf/10.1002/pra2.487},
year = {2021},
  note      = {\href{https://doi.org/10.1002/pra2.487}{doi:10.1002/pra2.487}}
}

@inproceedings{tankelevitch2024metacognitive,
author = {Tankelevitch, Lev and Kewenig, Viktor and Simkute, Auste and Scott, Ava Elizabeth and Sarkar, Advait and Sellen, Abigail and Rintel, Sean},
title = {The Metacognitive Demands and Opportunities of Generative AI},
year = {2024},
isbn = {9798400703300},
publisher = {Association for Computing Machinery},
address = {New York, NY, USA},
url = {https://doi.org/10.1145/3613904.3642902},
doi = {10.1145/3613904.3642902},
booktitle = {Proceedings of the 2024 CHI Conference on Human Factors in Computing Systems},
articleno = {680},
numpages = {24},
location = {Honolulu, HI, USA},
series = {CHI '24},
  note      = {\href{https://doi.org/10.1145/3613904.3642902}{doi:10.1145/3613904.3642902}}
}

@article{wang2023measuring,
author = {Bingcheng Wang and Pei-Luen Patrick Rau and Tianyi Yuan},
title = {Measuring user competence in using artificial intelligence: validity and reliability of artificial intelligence literacy scale},
journal = {Behaviour \& Information Technology},
volume = {42},
number = {9},
pages = {1324--1337},
year = {2023},
publisher = {Taylor \& Francis},
doi = {10.1080/0144929X.2022.2072768},
URL = { https://doi.org/10.1080/0144929X.2022.2072768},
eprint = { https://doi.org/10.1080/0144929X.2022.2072768},
  note      = {\href{https://doi.org/10.1080/0144929X.2022.2072768}{doi:10.1080/0144929X.2022.2072768}}
}

@misc{drosos2025itmakesthinkprovocations,
      title={"It makes you think": Provocations Help Restore Critical Thinking to AI-Assisted Knowledge Work}, 
      author={Ian Drosos and Advait Sarkar and Xiaotong and Xu and Neil Toronto},
      year={2025},
      eprint={2501.17247},
      archivePrefix={arXiv},
      primaryClass={cs.HC},
      url={https://arxiv.org/abs/2501.17247},
      note      = {\href{https://doi.org/10.48550/arXiv.2501.17247}{doi:10.48550/arXiv.2501.17247}}
}

@Article{gerlich2025b,
AUTHOR = {Gerlich, Michael},
TITLE = {From Offloading to Engagement: An Experimental Study on Structured Prompting and Critical Reasoning with Generative AI},
JOURNAL = {Data},
VOLUME = {10},
YEAR = {2025},
NUMBER = {11},
ARTICLE-NUMBER = {172},
URL = {https://www.mdpi.com/2306-5729/10/11/172},
ISSN = {2306-5729},
DOI = {10.3390/data10110172},
  note      = {\href{https://doi.org/10.3390/data10110172}{doi:10.3390/data10110172}}
}

@misc{holzner2025generative,
      title={Generative AI and Creativity: A Systematic Literature Review and Meta-Analysis}, 
      author={Niklas Holzner and Sebastian Maier and Stefan Feuerriegel},
      year={2025},
      eprint={2505.17241},
      archivePrefix={arXiv},
      primaryClass={cs.HC},
      url={https://arxiv.org/abs/2505.17241},
      note      = {\href{https://doi.org/10.48550/arXiv.2505.17241}{doi:10.48550/arXiv.2505.17241}}
}

@article{
jakesch2023human,
author = {Maurice Jakesch  and Jeffrey T. Hancock  and Mor Naaman },
title = {Human heuristics for AI-generated language are flawed},
journal = {Proceedings of the National Academy of Sciences},
volume = {120},
number = {11},
pages = {e2208839120},
year = {2023},
doi = {10.1073/pnas.2208839120},
URL = {https://www.pnas.org/doi/abs/10.1073/pnas.2208839120},
eprint = {https://www.pnas.org/doi/pdf/10.1073/pnas.2208839120},
  note      = {\href{https://doi.org/10.1073/pnas.2208839120}{doi:10.1073/pnas.2208839120}}
}

@inproceedings{kumar2024creativity,
author = {Kumar, Harsh and Vincentius, Jonathan and Jordan, Ewan and Anderson, Ashton},
title = {Human Creativity in the Age of LLMs: Randomized Experiments on Divergent and Convergent Thinking},
year = {2025},
isbn = {9798400713941},
publisher = {Association for Computing Machinery},
address = {New York, NY, USA},
url = {https://doi.org/10.1145/3706598.3714198},
doi = {10.1145/3706598.3714198},
booktitle = {Proceedings of the 2025 CHI Conference on Human Factors in Computing Systems},
articleno = {23},
numpages = {18},
location = {
},
series = {CHI '25},
  note      = {\href{https://doi.org/10.1145/3706598.3714198}{doi:10.1145/3706598.3714198}}
}

@article{wan2025diverse,
title = {Diverse AI personas can mitigate the homogenization effect in human-AI collaborative ideation},
journal = {Computers in Human Behavior: Artificial Humans},
volume = {8},
pages = {100289},
year = {2026},
issn = {2949-8821},
doi = {10.1016/j.chbah.2026.100289},
url = {https://www.sciencedirect.com/science/article/pii/S294988212600040X},
author = {Yun Wan and Yoram M. Kalman},
  note      = {\href{https://doi.org/10.1016/j.chbah.2026.100289}{doi:10.1016/j.chbah.2026.100289}}
}

@book{guilford1967nature,
  author    = {Guilford, Joy Paul},
  title     = {The Nature of Human Intelligence},
  publisher = {McGraw-Hill. },
  year      = {1967},
  url       = {https://archive.org/details/natureofhumanint0000guil},
  note      = {\href{https://archive.org/details/natureofhumanint0000guil}{archive.org url:natureofhumanint0000guil}}
}

@article{silvia2008assessing,
  author    = {Silvia, Paul J. and Winterstein, Beate P. and Willse, John T. and Barona, Christopher M. and Cram, Joshua T. and Hess, Karl I. and Martinez, Jenna L. and Richard, Crystal A.},
  title     = {Assessing Creativity with Divergent Thinking Tasks: Exploring the Reliability and Validity of New Subjective Scoring Methods},
  journal   = {Psychology of Aesthetics, Creativity, and the Arts},
  volume    = {2},
  number    = {2},
  pages     = {68--85},
  year      = {2008},
  doi       = {10.1037/1931-3896.2.2.68},
  url       = {https://doi.org/10.1037/1931-3896.2.2.68},
  note      = {\href{https://doi.org/10.1037/1931-3896.2.2.68}{doi:10.1037/1931-3896.2.2.68}}
}

@article{schepman2020gaais,
title = {Initial validation of the general attitudes towards Artificial Intelligence Scale},
journal = {Computers in Human Behavior Reports},
volume = {1},
pages = {100014},
year = {2020},
issn = {2451-9588},
doi = {10.1016/j.chbr.2020.100014},
url = {https://www.sciencedirect.com/science/article/pii/S2451958820300142},
author = {Astrid Schepman and Paul Rodway},
  note      = {\href{https://doi.org/10.1016/j.chbr.2020.100014}{doi:10.1016/j.chbr.2020.100014}}
}

@misc{padmakumar2026offloading,
      title={Offloading Score: Measuring AI Reliance Through Counterfactual Workflows}, 
      author={Vishakh Padmakumar and Lujain Ibrahim and Zora Zhiruo Wang and Jennifer Wang and Q. Vera Liao and Diyi Yang},
      year={2026},
      eprint={2605.29392},
      archivePrefix={arXiv},
      primaryClass={cs.SE},
      url={https://arxiv.org/abs/2605.29392},
      note      = {\href{https://doi.org/10.48550/arXiv.2605.29392}{doi:10.48550/arXiv.2605.29392}}
}

@ARTICLE{wu2026aidep,
    
AUTHOR={Wu, Houyu  and Ni, Haiyang  and Luo, Wenfu  and Wu, Tenglong },
           
TITLE={Development and validation of the AI dependence scale for Chinese undergraduates and a preliminary exploration},
          
JOURNAL={Frontiers in Psychology},
          
VOLUME={16},
  
YEAR={2026},
  
URL={https://www.frontiersin.org/journals/psychology/articles/10.3389/fpsyg.2025.1725393},
  
DOI={10.3389/fpsyg.2025.1725393},
  
ISSN={1664-1078},

note      = {\href{https://doi.org/10.3389/fpsyg.2025.1725393}{doi:10.3389/fpsyg.2025.1725393}}}

@article{garciacastro2025said,
  title   = {Artificial intelligence dependence in academic tasks: Design and validation of the {SAID} questionnaire},
  author  = {Garcia Castro, Raul Alberto and Bartesaghi Aste, William Maximo and Morales Quezada, Jose Luis and Arocutipa Huanacuni, Lupita Esmeralda},
  journal = {Online Journal of Communication and Media Technologies},
  volume  = {15},
  number  = {4},
  pages   = {e202529},
  year    = {2025},
  doi     = {10.30935/ojcmt/17303},
  note      = {\href{https://doi.org/10.30935/ojcmt/17303}{doi:10.30935/ojcmt/17303}},
}

@misc{jin2024glat,
      title={GLAT: The Generative AI Literacy Assessment Test}, 
      author={Yueqiao Jin and Roberto Martinez-Maldonado and Dragan Gašević and Lixiang Yan},
      year={2024},
      eprint={2411.00283},
      archivePrefix={arXiv},
      primaryClass={cs.HC},
      url={https://arxiv.org/abs/2411.00283},
      note      = {\href{https://doi.org/10.48550/arXiv.2411.00283}{doi:10.48550/arXiv.2411.00283}},
}

@article{shaw2026surrender,
  title   = {Thinking---Fast, Slow, and Artificial: How {AI} is Reshaping Human Reasoning and the Rise of Cognitive Surrender},
  author  = {Steven D. Shaw and Gideon Nave},
  journal = {The Wharton School Research Paper},
  year    = {2026},
  doi     = {10.2139/ssrn.6097646},
  note      = {\href{https://doi.org/10.2139/ssrn.6097646}{doi:10.2139/ssrn.6097646}},
}

@misc{liu2026persistence,
      title={AI Assistance Reduces Persistence and Hurts Independent Performance}, 
      author={Grace Liu and Brian Christian and Tsvetomira Dumbalska and Michiel A. Bakker and Rachit Dubey},
      year={2026},
      eprint={2604.04721},
      archivePrefix={arXiv},
      primaryClass={cs.AI},
      url={https://arxiv.org/abs/2604.04721},
      note      = {\href{https://doi.org/10.48550/arXiv.2604.04721}{doi:10.48550/arXiv.2604.04721}},
}

@misc{rismanchian2026faster,
      title={Faster Completion, Less Learning: Generative AI Reduced Study Time on Math Problems and the Knowledge They Build}, 
      author={Sina Rismanchian and Hasan Uzun and Jeffrey Matayoshi and Eric Cosyn and Eyad Kurd-Misto},
      year={2026},
      eprint={2605.21629},
      archivePrefix={arXiv},
      primaryClass={cs.CY},
      url={https://arxiv.org/abs/2605.21629},
      note      = {\href{https://doi.org/10.48550/arXiv.2605.21629}{doi:10.48550/arXiv.2605.21629}},
}

@article{hong2025cognitive, 

title={Cognitive Offload Instruction with Generative AI: A Quasi‑Experimental Study on Critical Thinking Gains in English Writing}, 

volume={7}, 

url={https://journals.bilpubgroup.com/index.php/fls/article/view/10072}, 

DOI={10.30564/fls.v7i7.10072}, 

abstractNote={&amp;lt;p&amp;gt;This study explores the impact of generative AI-enabled cognitive offload instruction on the development of critical thinking skills in English essay writing among first-year university students. A quasi-experimental design was employed, comparing traditional instruction with an AI-augmented pedagogy that delegated lower-order writing tasks to generative AI tools, allowing students to focus on analysis, evaluation, and reflection. Over 12 weeks, 240 participants engaged in structured writing cycles involving AI brainstorming, individual critique, peer-AI co-revision, and reflective journaling. Results revealed that the AI-enabled cognitive offload group demonstrated significantly greater improvements in standardized critical thinking assessments and produced higher-quality essays in terms of logical coherence, evidence use, and originality. Mediation analysis indicated that cognitive offloading behavior partially explained the relationship between AI use and critical thinking gains. The findings suggest that when generative AI is integrated into pedagogy through deliberate scaffolding, it can enhance rather than hinder higher-order thinking. This study highlights the importance of balancing technological efficiency with instructional strategies that promote active engagement, metacognitive reflection, and collaborative learning. It offers practical implications for educators seeking to incorporate AI tools without compromising the development of essential cognitive skills, proposing that structured cognitive offload instruction can serve as an effective approach to fostering critical thinking in second-language writing contexts.&amp;lt;/p&amp;gt;}, 

number={7}, journal={Forum for Linguistic Studies}, 

author={Hong, Hui and Vate-U-Lan, Poonsri and Viriyavejakul, Chantana}, 

year={2025}, 

month={Jul.}, 

pages={325–334},

note      = {\href{https://doi.org/10.30564/fls.v7i7.10072}{doi:10.30564/fls.v7i7.10072}},

}

\end{document}